\documentclass[conference]{IEEEtran}
\IEEEoverridecommandlockouts
\usepackage{amsmath,amsfonts}
\usepackage{algorithmic}
\usepackage{algorithm}
\usepackage{array}
\usepackage[caption=false,font=normalsize,labelfont=sf,textfont=sf]{subfig}
\usepackage{textcomp}
\usepackage{stfloats}
\usepackage{url}
\usepackage{verbatim}
\usepackage{graphicx}
\usepackage{cite}
\usepackage{adjustbox}
\def\BibTeX{{\rm B\kern-.05em{\sc i\kern-.025em b}\kern-.08em
    T\kern-.1667em\lower.7ex\hbox{E}\kern-.125emX}}

\usepackage{booktabs}
\usepackage{rotating}

\begin{document}

\title{A Deep Learning Model for Predicting Transformation Legality}

\author{
\IEEEauthorblockN{Avani Tiwari}
\IEEEauthorblockA{
\textit{Department of Computer Science} \\
\textit{New York University Abu Dhabi} \\
Abu Dhabi, UAE \\
at4535@nyu.edu}
\and
\IEEEauthorblockN{Yacine Hakimi}
\IEEEauthorblockA{
\textit{Laboratoire de Méthodes de Conception de Systèmes} \\
\textit{Ecole nationale Supérieure d'Informatique} \\
Oued Smar, BP M68, 16309, Algiers, Algeria \\
y\_hakimi@esi.dz}
\and
\IEEEauthorblockN{Riyadh Baghdadi}
\IEEEauthorblockA{
\textit{Department of Computer Science} \\
\textit{New York University Abu Dhabi} \\
Abu Dhabi, UAE \\
baghdadi@nyu.edu}
}

\maketitle

\begin{abstract}

Compilers must check the legality of code transformations to guarantee the correctness of applying a sequence of code transformations to a given code. While such a legality check needs to be precisely computed in general, we can use an approximate legality prediction model in certain cases, such as training a reinforcement learning (RL) agent for schedule prediction.
In this paper, we propose an approximate method for legality checks. We propose a novel DL model for predicting the legality of transformations.
The model takes the code representation and a list of transformations as input and predicts whether applying those transformations to the code is legal.
We implement and evaluate the proposed model, demonstrating its effectiveness. Our evaluation shows an F1 score of 0.91 on a test set of randomly generated programs. To further evaluate the model in a practical scenario, we used the model to replace the legality check used during the training of an RL agent designed for automatic code optimization. We demonstrate that such a replacement enables the agent to train on twice as many steps, resulting in faster training and reducing resource usage by approximately 80\% for CPU and 35\% for RAM. The agent trained using this approach maintains comparable performance, with only a 4\% reduction on benchmarks from the Polybench suite compared to the traditional method.

\end{abstract}


\section{Introduction}
Machine learning workloads with billions of parameters demand aggressive compiler optimizations to reduce training and inference times. Code optimization is necessary to generate efficient code, thereby reducing the inference and training times of these models. Enabling compilers to optimize code automatically is a promising solution to this problem.

The problem of automatic code optimization can be modeled as a search problem, where the goal is to find the sequence of code transformations (code optimizations) that minimizes the program's execution time. The compiler uses a search space exploration method to explore the search space of code transformations. Each potential solution in this search space is a sequence of code transformations (code optimizations). For each possible solution explored in the search space, the compiler must verify whether it is legal (i.e., whether applying the sequence of code transformations preserves the original semantics of the code).

The legality of a sequence of transformations is checked by performing a dependency analysis and then applying a legality check \cite{feautrier_array_1988,violateddep}. This legality check is computationally expensive and time-consuming, especially when exploring a large search space. Two statements are said to be dependent when they access the exact memory location, and one has a write access. In such a case, their original execution order has to be preserved~\cite{scheduling}. While a legality check is usually necessary to guarantee the correctness of code transformations applied, there are a few scenarios where a precise legality check is not required. The legality is said to be approximate in this case. Such an approximate legality check is useful, for example, if the user wants to accelerate the legality check at the expense of its precision. Such a use case motivates the proposal of a DL model for an approximate but fast legality check. Such a model has the advantage of being differentiable, which opens the door to many applications that would be hard otherwise.

One scenario where an approximate legality check is useful is the development of an RL agent for automatic code optimization in a polyhedral compiler. Performing legality checks in a polyhedral compiler is time-consuming (due to the use of algorithms with exponential complexity). In our experiments, legality checks take one-third of the training time of an RL agent. This is an obstacle that needs to be solved to accelerate the training of the RL agent. This is important because training an RL agent takes weeks. It requires trying tens of millions of actions (where each action is a code transformation) and verifying the legality of each action. Replacing the classical legality check with an approximate one (only during training) could significantly accelerate the training while possibly having a minor negative effect on the quality of the final RL agent. Note that the final RL agent, when deployed, does not require an approximate legality check. It will rather use a precise one to ensure the correctness of the selected transformations. 

Another scenario where an approximate legality check is useful is when using gradient-based optimization to search for the best code transformations. In such a scenario, one can leverage a DL-based cost model (i.e., that predicts a performance metric) to search for the best optimizations. As this model is differentiable, it can be used with a gradient-based optimization method to explore the optimal transformations. The issue here is that cost models do not consider the legality of transformations. Naively using this model along with a gradient-based search will result in illegal transformations. Developing a DL model to predict the legality would help address this issue. One can use a gradient-based search to minimize an objective function that combines the cost of a given transformation and its legality. At the end of the process, we must conduct a precise dependency analysis to verify the correctness of the final result. However, the likelihood of finding a valid and effective transformation is greater.

This paper proposes a DL model to predict the legality of applying a sequence of loop transformations on a given code. This model takes as input a code representation and a sequence of transformations and predicts whether it is legal to apply the transformations to the code. Using this model accelerates the training of an RL agent and reduces resource usage by approximately 80\% for CPU and 35\% for RAM.

This model has the advantage of being fast and differentiable (compared to classical dependency analysis). Because it is approximate, it is well-suited to certain scenarios but not as a general replacement for classical dependency analysis.

The contributions of the paper are as follows:
\begin{itemize}
\item We introduce the first DL-based classification model to predict the legality of applying a sequence of code transformations on a given code. 
\item We train and evaluate the proposed model, demonstrating its effectiveness in reducing resource usage and accelerating RL training with minimal performance trade-offs.
\item We release and open-source our model and dataset used to train the model.
\end{itemize}

\section{Proposed Model} 
The proposed model architecture is mainly inspired by the one proposed by \cite{merouani2024looper}. Since that work uses a deep learning (DL) model for performance prediction, we propose to perform the necessary adaptations to the model architecture and training policy to accommodate the task of legality prediction. In \cite{merouani2024looper}, legality was considered during exploration, and a traditional legality check using dependency analysis was used when finding the best schedules. The model in \cite{merouani2024looper} is not designed to predict legality, which is the focus of our proposed work.


\begin{figure} [t]%
    \centering
    \includegraphics[width=0.45\textwidth]{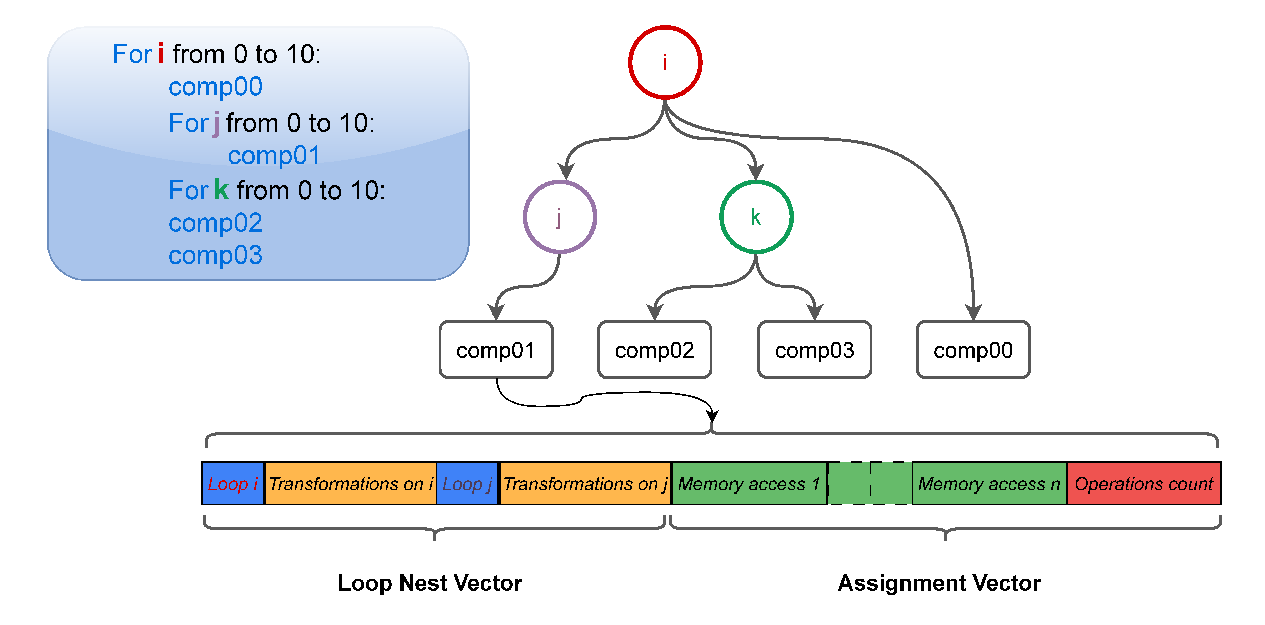} 
    \caption{Illustration of Program Representation with Loop and Assignment Vectors }
    \label{Program_representation}
\end{figure}

\begin{figure}[t] %
    \centering
    \includegraphics[width=0.45\textwidth]{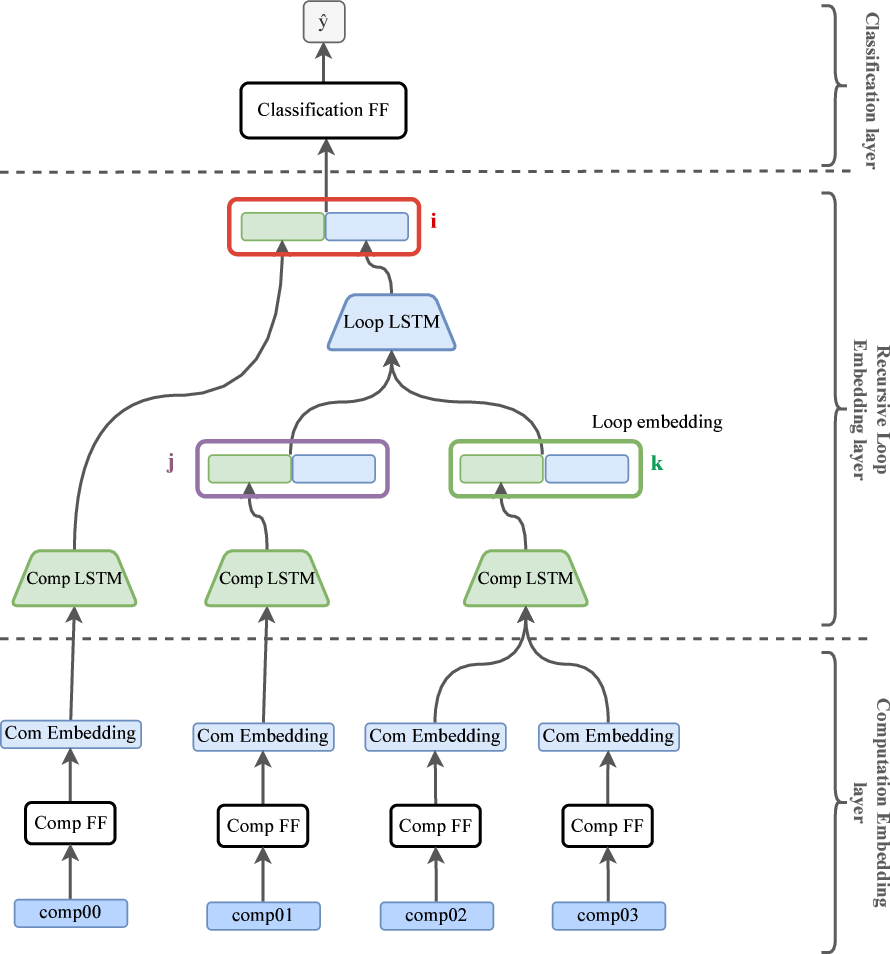} 
    \caption{ DL Model Architecture for Transformation Legality Prediction }
    \label{model_architecture}
\end{figure}


\subsection{Program Characterization}

The programs used for training the models are represented by extracting high-level information and storing it in a compact variable-size representation, the same representation as the TIRAMISU cost model\cite{merouani2024looper}.

We use the same abstract program representation as the Tiramisu cost model \cite{merouani2024looper}, encoding loop nests, applied transformations, and memory access patterns \ref{Program_representation}.

\subsubsection{Loop-nest Representation}
The loop-nest structure is represented by capturing the extent of each loop level surrounding computations. Each loop representation is followed by a representation of the transformations applied at this level.

\subsubsection{Loop transformation Representation}

The representation of each transformation involves encoding its type and parameters. After each loop level representation, we add a series of Boolean indicators denoting whether particular transformations have been applied, followed by the corresponding transformation parameters.

\subsubsection{Assignments representation}
The assignment representation includes the memory access patterns of each accessed buffer and its ID, followed by the count of each arithmetic operation performed in the assignment. Memory accesses are stored in affine matrices, similar to the polyhedral model~\cite{feautrier2011polyhedron}, where k rows (the number of dimensions of the access buffer) and n + 1 columns (where n is the loop depth).

\subsubsection{Program Structure Representation}
The overall program structure is a hierarchical tree, with computations as leaf nodes and loop levels as internal nodes. This tree format enables the recursive processing of loop nests, allowing the model to incorporate structural information efficiently. The tree’s organization reflects nesting relationships and the iteration space of each computation, providing a foundation for assessing transformation legality by the model.

\subsection{Model Architecture}

We address the task of predicting the legality of a transformation schedule as a binary classification problem. Our model, based on an architecture similar to that proposed by ~\cite{merouani2024looper}, with an output of 1 if the schedule is legal and 0 otherwise. 

The model consists of 3 layers as shown in Figure \ref{model_architecture}: a computation embedding layer, a recursive loop embedding layer, and a classification layer. 

\subsubsection{Computation Embedding Layer}

The computation embedding layer transforms non-boolean features of the program's computation vectors by applying a logarithmic transformation to scale feature values. These transformed features are then processed through a feedforward neural network (FNN), which generates embeddings for each computation. This embedding process condenses the information into a format suitable for downstream layers, ensuring that each computation’s characteristics are encoded effectively. 

\subsubsection{Recursive Loop Embedding Layer}
The recursive loop embedding layer operates hierarchically, following the program tree structure. This layer incorporates two Long Short-Term Memory (LSTM) cells at each loop level: one processes embeddings of computations directly nested within the loop level, while the other processes embeddings from previous loop levels, capturing dependencies across nested structures. The two LSTM outputs are combined to generate a loop-level embedding representation. The model constructs a comprehensive program embedding that considers the positional and hierarchical relationships among computations by recursively applying this structure across loop levels.

\subsubsection{Classification Layer}

The classification layer takes the program embedding vector generated by the previous layer and processes it through an FNN, followed by a sigmoid activation function. We use 0.5 as a threshold to predict legality, which was selected through iterative experimentation for optimizing the model's predictive accuracy.

We employ Binary Cross-Entropy as the loss function to train the model, which measures the divergence between the predicted and actual legality classifications. Model optimization is performed using the AdamW optimizer \cite{loshchilov2017decoupled}, which combines adaptive learning rates with decoupled weight decay to improve convergence and prevent overfitting.

\subsection{Data Set\label{dataset}}\

We collect a dataset of legal and illegal sequences of transformation on synthetic TIRAMISU programs. 

The synthetic programs are generated using the program generator developed in \cite{tiramisu21}. This generator is designed to generate programs that are representative of real programs. The input data is generated automatically with a random size. Since the generated programs are data-independent, the actual data values are not important. 

 We run an RL agent on the generated synthetic programs to get the transformation sequences. This agent explores the search space by applying code transformations on the programs while checking the legality of each candidate sequence of transformations using dependency analysis. The transformations explored by the agent are loop interchange, loop skewing, loop reversal, loop parallelization, loop tiling, and loop unrolling. For each synthetic program, we record both legal and illegal transformation sequences that the RL agent encounters. This collection of synthetic programs, transformations, and their legality will be used as a training dataset. For some programs, the legal transformations had to be removed to balance the legal/illegal ratio of the dataset.

The training data set has 5.8M points, 3.2M legal points, and 2.6M illegal points. The testing data set has 1.21M points, 0.64M legal points, and 0.57M illegal points.

\section{Evaluation}

We evaluate the accuracy of our legality prediction model on the test set. We also evaluate the impact of using our model to train an RL agent to auto-optimize programs. We compare three RL agents: one trained using the classical legality check used in Tiramisu, the second trained using our legality prediction model, and the third trained using a random legality model. The results are described in \ref{rl_evaluation}.

\subsection{Evaluation Metric}
To evaluate the model's accuracy, we will use the F1 score. The F1 score is the harmonic mean of precision and recall. It provides a quantitative analysis of the model's performance.

$F1 = \frac{2 \times (Precision \times Recall)}{Precision + Recall} \\$

$Precision = \frac{\text{True\_Positives}}{\text{True\_Positives} + \text{False\_Positives}} \\$

$Recall = \frac{\text{True\_Positives}}{\text{True\_Positives} + \text{False\_Negatives}}\\$

\subsection{The Performance on the Test Data}

The precision on the randomly generated dataset is 0.92, the recall is 0.90, and the F1 score is 0.91. 

\begin{figure}[t] %
    \centering
    \includegraphics[width=0.4\textwidth]{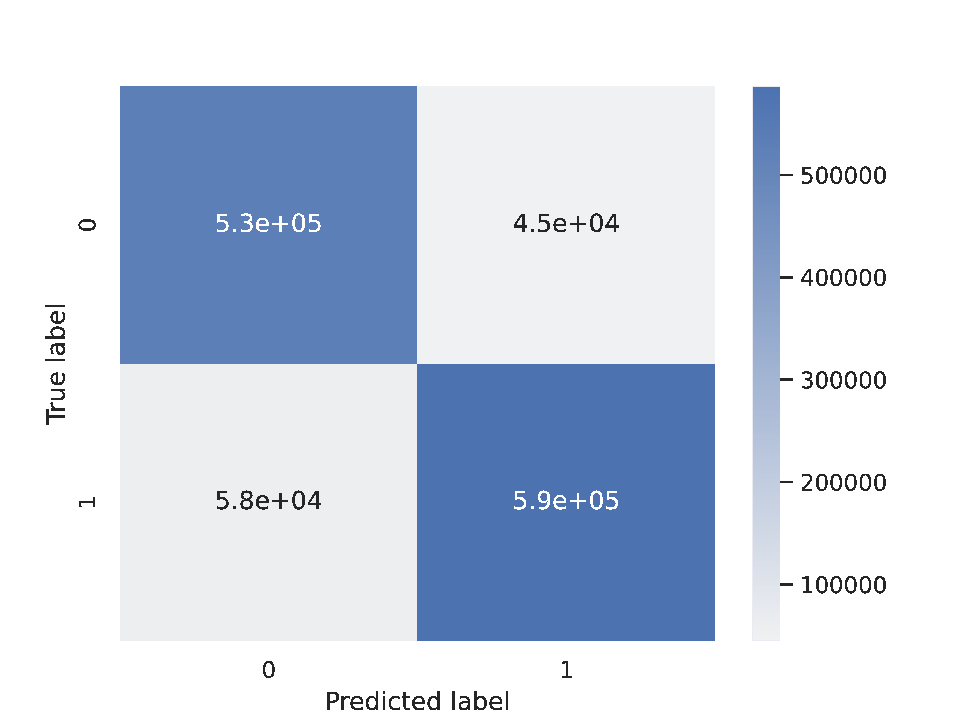}
    \caption{Confusion matrix for the randomly generated test data}
    \label{fig:be}
\end{figure}

\begin{figure*} %
    \centering
    \includegraphics[width=0.6\textwidth]{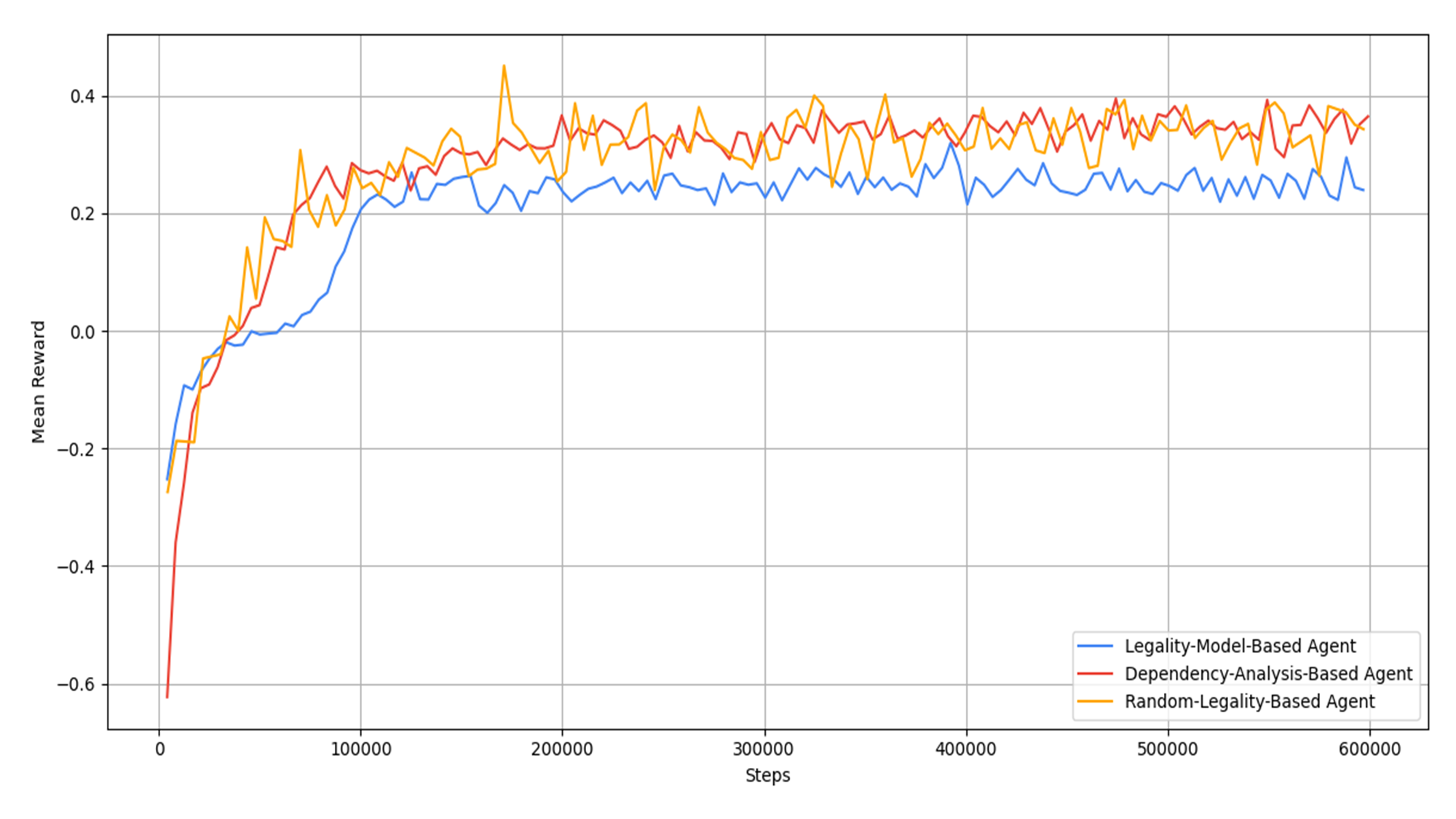} 
    \caption{Comparison of mean reward between the Legality-Model-Based Agent, the Dependency-Analysis-Based Agent,and Random-Legality-Based Agent over training steps. }
    \label{fig:mean_reward}
\end{figure*}

\begin{figure*}
    \centering
    \includegraphics[width=0.8\textwidth]{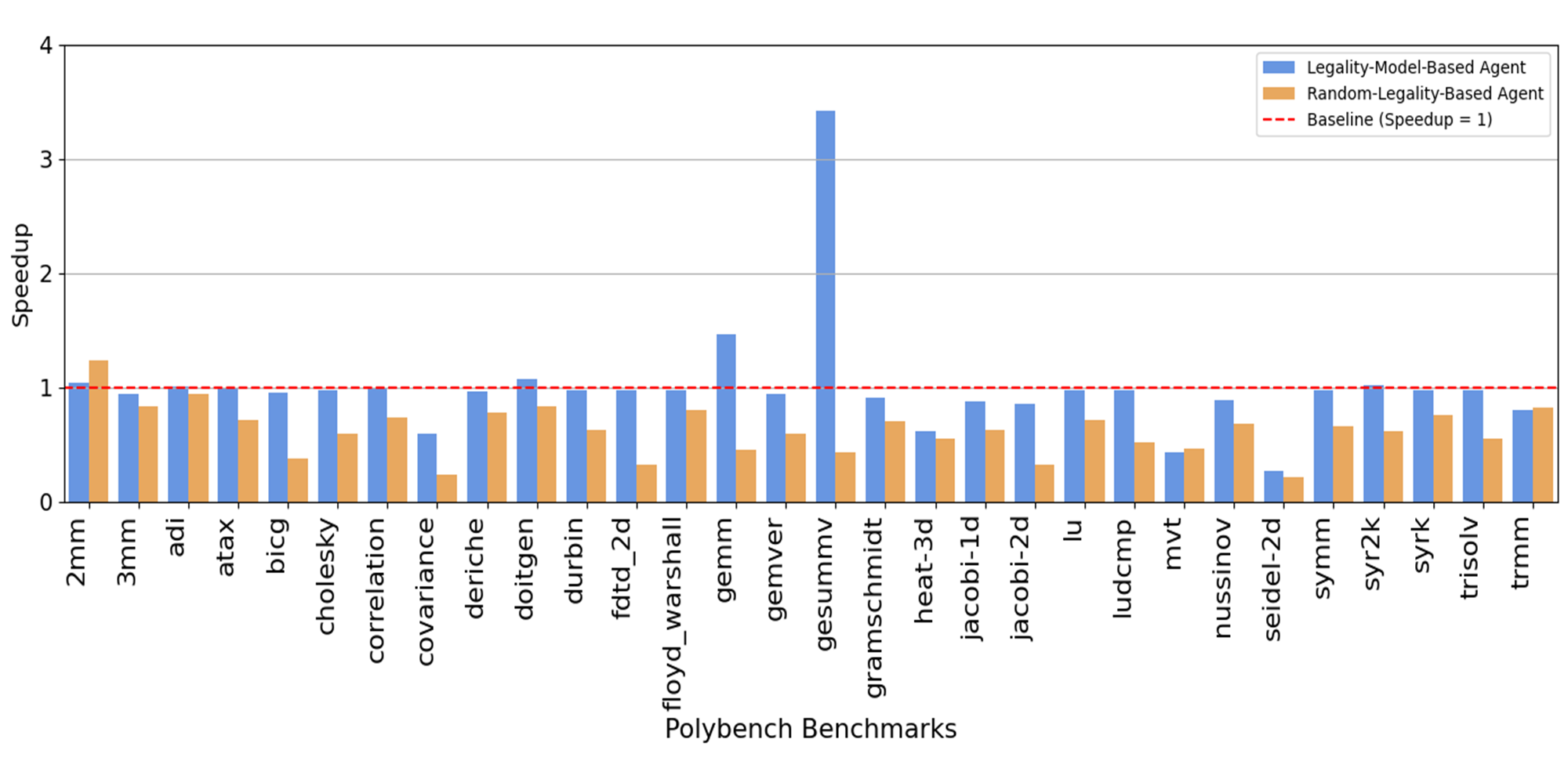} 
    \caption{Geometric means of the speedup of the Legality-Model-Based Agent and Random-Legality-Based Agent over the dependency-analysis agent on the Polybench suite }
    \label{fig:rl_poly}
\end{figure*}

\begin{figure} %
    \centering
    \includegraphics[width=0.5\textwidth]{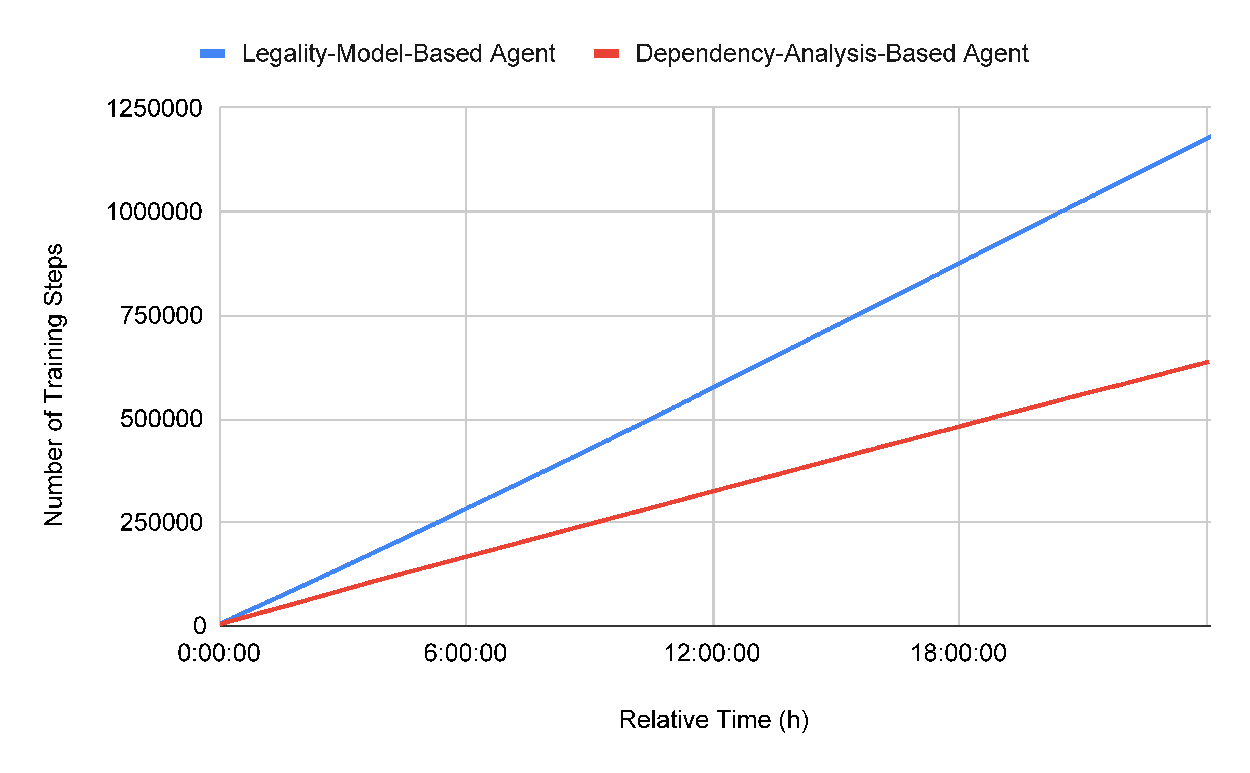} 
    \caption{Evolution of the number of training steps through time for the model-based agent and dependency-analysis-based one through 24 hours of training}
    \label{fig:rl_time}
\end{figure}

\begin{figure*} %
    \centering
    \includegraphics[width=0.55\textwidth]{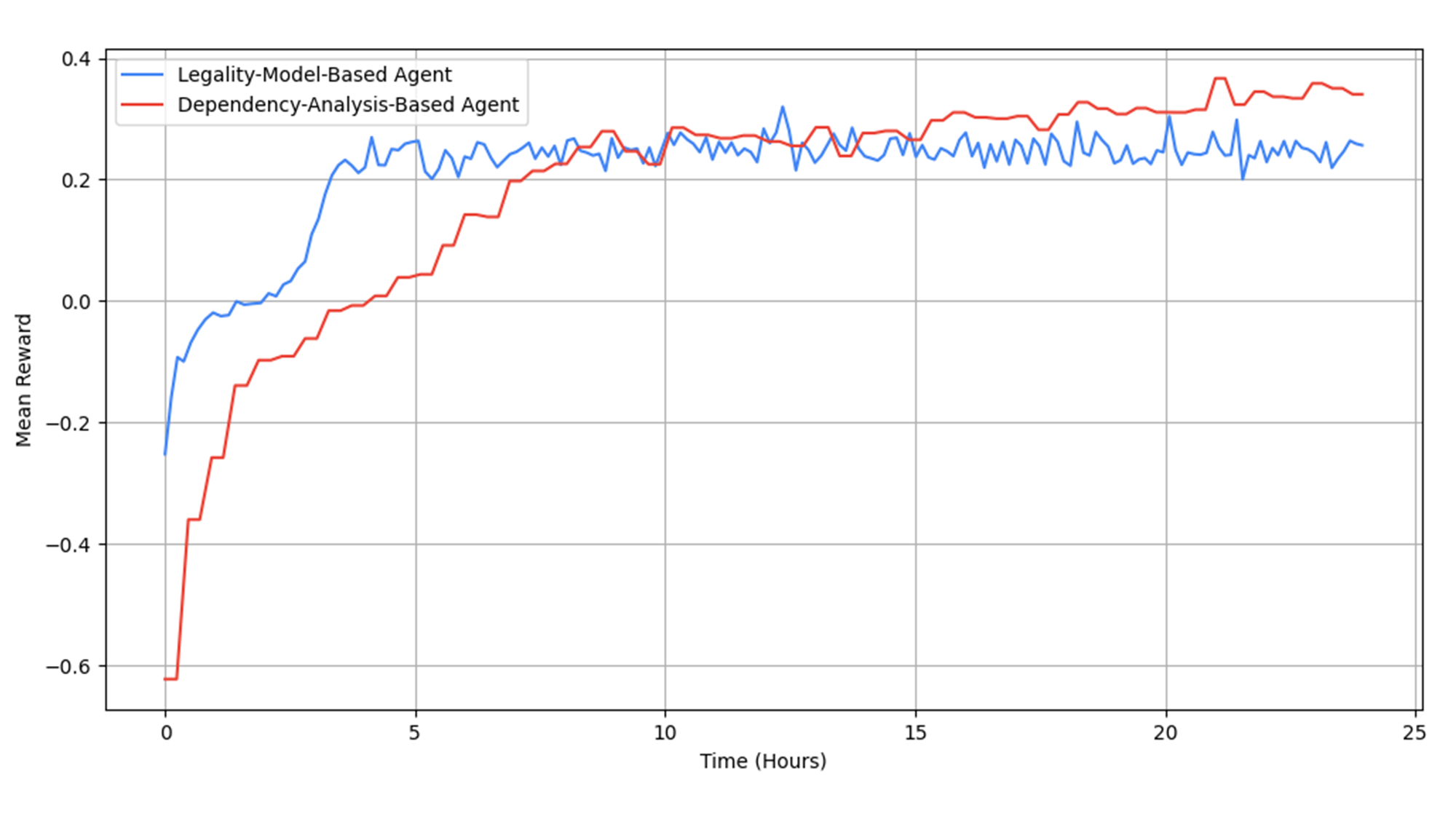} 
    \caption{Comparison of mean reward between the Legality-Model-Based Agent and the Dependency-Analysis-Based Agent over time.}
    \label{fig:Mean_reward_time}
\end{figure*}

Precision is the ratio of true positive predictions to the total positive predictions. A precision equal to 1 means a perfect model. In our case, we obtained a precision of 0.92. That means that 92\% of the total labels our model classifies as positive are indeed positive (legal).

Recall is the ratio of true positive predictions to the actual positives in the data. In our case, our model obtains a recall score of 0.90. That is, among all the actual positive data points (legal data schedules), our model correctly predicted 90\%.

The F1 score is the harmonic mean of precision and recall. A high F1 score (close to 1) indicates both high precision and recall, suggesting that the model effectively identifies positive cases and is accurate in its positive predictions.
A low F1 score (close to 0) indicates that either precision, recall, or both are low. This means the model is either not identifying positive cases correctly (low recall) or is making many incorrect positive predictions (low precision), or both. Our model obtained an F1 score of 0.91.

\subsection{Case Study: Accelerating the Training of a Code Optimization RL Agent}
\label{rl_evaluation}

Our proposed legality check model is, on average, an order of magnitude faster than the classical legality check. Generally, the proposed DL model has a linear execution time, while the polyhedral legality check algorithms have an exponential complexity. This makes the proposed model an interesting alternative in scenarios where checking legality is computationally expensive but not strictly essential.

We have been developing in parallel an RL agent for automatic code optimization. Through that project, we identified that approximately 30\% of the RL agent's training time is spent performing legality checks, significantly prolonging the training period (often taking weeks), as each decision requires recompiling and running code to assess the reward.

This bottleneck prompted us to hypothesize that training the RL agent with some illegal transformations might not significantly degrade the schedules' quality produced during deployment, as the policy network's primary goal is to optimize schedule quality, not correctness. We assume that during deployment, classical legality checks will ensure the correctness of transformations, allowing the agent to focus on finding a near-optimal schedule.
On the other hand, we assume that training an RL agent to predict optimal schedules without regard for their correctness will lead to a poor policy, resulting in poor performance in program optimization. Additionally, this agent would frequently produce illegal schedules during deployment, increasing the time and computational resources required to discover correct and near-optimal schedules.

Consequently, to explore this trade-off, we propose using our model during training to reduce training time and computational resource usage while maintaining a comparable quality of schedules generated by the RL agent during deployment.

To test this hypothesis and evaluate the effectiveness of our model, we conducted a comparative test using a limited set of approximately 25,000 training programs. The purpose of this test was to directly compare the efficiency and performance of training between three agents, rather than create a fully optimized agent for deployment. This smaller-scale test allows for effective observation of differences in training dynamics without the computational expense of using an exhaustive dataset.

In this test, we trained three RL agents: one using the classical Tiramisu legality check (which relies on dependency analysis), another using our DL-based legality check model, a third one with a Random Legality Model, which predicts the legality of transformations with a 50-50 probability(This will provides a reference for how a policy trained with minimal regard for correctness affects performance).

The three agents trained for 24 hours and then evaluated on the Polybench benchmark suite~\cite{polybench}, widely used in the literature for polyhedral compilers evaluation. The benchmark comprises kernels collected from real applications, including diverse domains such as ML, linear algebra, and stencils.

    

We compared the geometric mean speedup for each benchmark with varying input sizes across the three agents.  Additionally, we also compared the number of illegal schedules generated by each agent during optimization to analyze how each approach influenced the efficiency of exploration.

\subsection{Mean Reward during the training} 
Figure \ref{fig:mean_reward} illustrates the Mean reward across training steps of the three agents: Dependency-Analysis-Based Agent (red), Legality-Model-Based Agent (blue), and Random-Legality-Based Agent (orange). The Random-Legality-Based agent reward is comparable to the dependency-analysis agent. However, the random agent's reward has higher fluctuations across training steps, with a larger variation from step to step than the other two agents. We believe that the randomness of legality prediction accounts for this variation. Such randomness not only lets it sometimes follow extremely profitable paths(Regardless of its legality), but it can also cause it to be highly penalized for illegal predicted transformations (the reward function penalizes illegal transformations to direct the agent to choose legal ones). The Legality-Model-Based Agent has lower rewards than the others. That would likely be due to prediction errors in the legality, valid schedules getting mistaken for illegal ones. Importantly, in this scenario, a higher reward in training does not mean a better agent, as the approximation of the legality-checking affects rewards for the Legality-Model and Random-Legality-Based agents. We would argue that testing the learned policies on benchmarks like Polybench may better indicate the actual performance.

\subsection{Speedup achieved on Polybench}
Figure \ref{fig:rl_poly} compares the performance of the three RL agents on the Polybench benchmark suite. The red line in the figure represents the baseline (Dependency-Analysis-Based Agent performance), the blue bars show the performance of the Legality-Model-Based Agent, and the orange bars show the Random-Legality-Based Agent performance. We can see that, except for six benchmarks, the Legality-Model-Based Agent performed quite similarly to the Dependency-Analysis-Based Agent on most benchmarks. On \texttt{covariance}, \texttt{heat-3d}, \texttt{seidel2d} and \texttt{mvt}, the Dependency-Analysis-Based agent can generate a much faster schedule. While it is the Legality-Model-Based Agent schedules that are faster for \texttt{gemm} and \texttt{gesummv}.
The Random-Legality-Based Agent performs much worse than the other two agents on most benchmarks. It achieves better speedup only on the \texttt{2mm} and comparable performance to the Legality-Model-Based Agent on \texttt{3mm},\texttt{adi}, and \texttt{trmm}. This indicates that a correct legality check model is crucial for the RL agent to explore legal high-performance schedules. 
In summary, the agent trained using the legality prediction model achieves a geometric mean of \textbf{0.96} compared to the baseline \textbf{1} of the agent trained with traditional Tiramisu dependency analysis, while the Random-Legality-Based Agent achieves a geometric mean of \textbf{0.62}.

\subsection{Trade-off Between Performance and Training Speed}
Using our model allows for some trade-off in accuracy for training speed. While the legality-model-based agent might occasionally misclassify some transformations, resulting in slightly lower rewards during training and reducing some performance on benchmarks, it completes the training process much faster. Figure \ref{fig:rl_time} shows that within the same 24-hour training period, the legality-model-based agent completed \textbf{1.182 million} training steps, nearly double the \textbf{0.638 million} steps completed by the dependency-analysis-based agent. In Figure \ref{fig:Mean_reward_time}, which illustrates the mean reward over time for both agents, we can see clearly that the Legality-Model-Based agent starts to converge earlier, achieving a stable mean reward approximately after 5 hours. In contrast, the Dependency-Analysis-Based Agent demonstrates a slower convergence, stabilizing around 9 hours after the training begins.

\begin{table}[t]
\centering
\caption{Comparison of the number of illegal transformations explored by each agent during the optimization process for each benchmark}
\resizebox{\columnwidth}{!}{
\label{tab:table1}
\begin{tabular}{lrrr}
\toprule
Benchmark & \parbox[t]{2cm}{Random-Legality-\\Based Agent} & \parbox[t]{2cm}{Legality-Model-\\Based Agent} & \parbox[t]{2cm}{Dependency-Analysis-\\Based Agent} \\
\midrule
2mm               & 50 & 14  & 9  \\
3mm               & 28  & 15  & 14  \\
adi               & 40  & 25  & 11  \\
atax              & 28  & 25  & 22  \\
bicg              & 30  & 18  & 19  \\
cholesky          & 76  & 58  & 47  \\
correlation       & 16  & 20  & 13  \\
covariance        & 21  & 29  & 22  \\
deriche           & 50  & 42  & 33  \\
doitgen           & 29  & 32  & 14  \\
durbin            & 42  & 31  & 29  \\
fdtd\_2d          & 39  & 25  & 19  \\
floyd\_warshall   & 18  & 13  & 9  \\
gemm              & 23  & 19  & 11  \\
gemver            & 41  & 24  & 17  \\
gesummv           & 43  & 7  & 8  \\
gramschmidt       & 82  & 56  & 41  \\
heat3d            & 33  & 27  & 22  \\
jacobi1d          & 30  & 19  & 12  \\
jacobi2d          & 32  & 12  & 6  \\
lu                & 42  & 42  & 39  \\
ludcmp            & 67  & 56  & 57  \\
mvt               & 47  & 32  & 23  \\
nussinov          & 60  & 43  & 21  \\
seidel2d          & 29  & 20  & 23  \\
symm              & 61  & 37  & 40  \\
syr2k             & 53  & 46  & 41  \\
syrk              & 64  & 37  & 33  \\
trisolv           & 37  & 25  & 23  \\
trmm              & 52  & 35  & 30  \\
\midrule
Sum               & 1263 & 884  & 708 \\
\bottomrule
\end{tabular}
}
\label{tab:benchmark_results}
\end{table}

\begin{figure*}[t]
	\centering
	\begin{minipage}[t]{0.48\textwidth}
	    \centering
        \includegraphics[width=\textwidth]{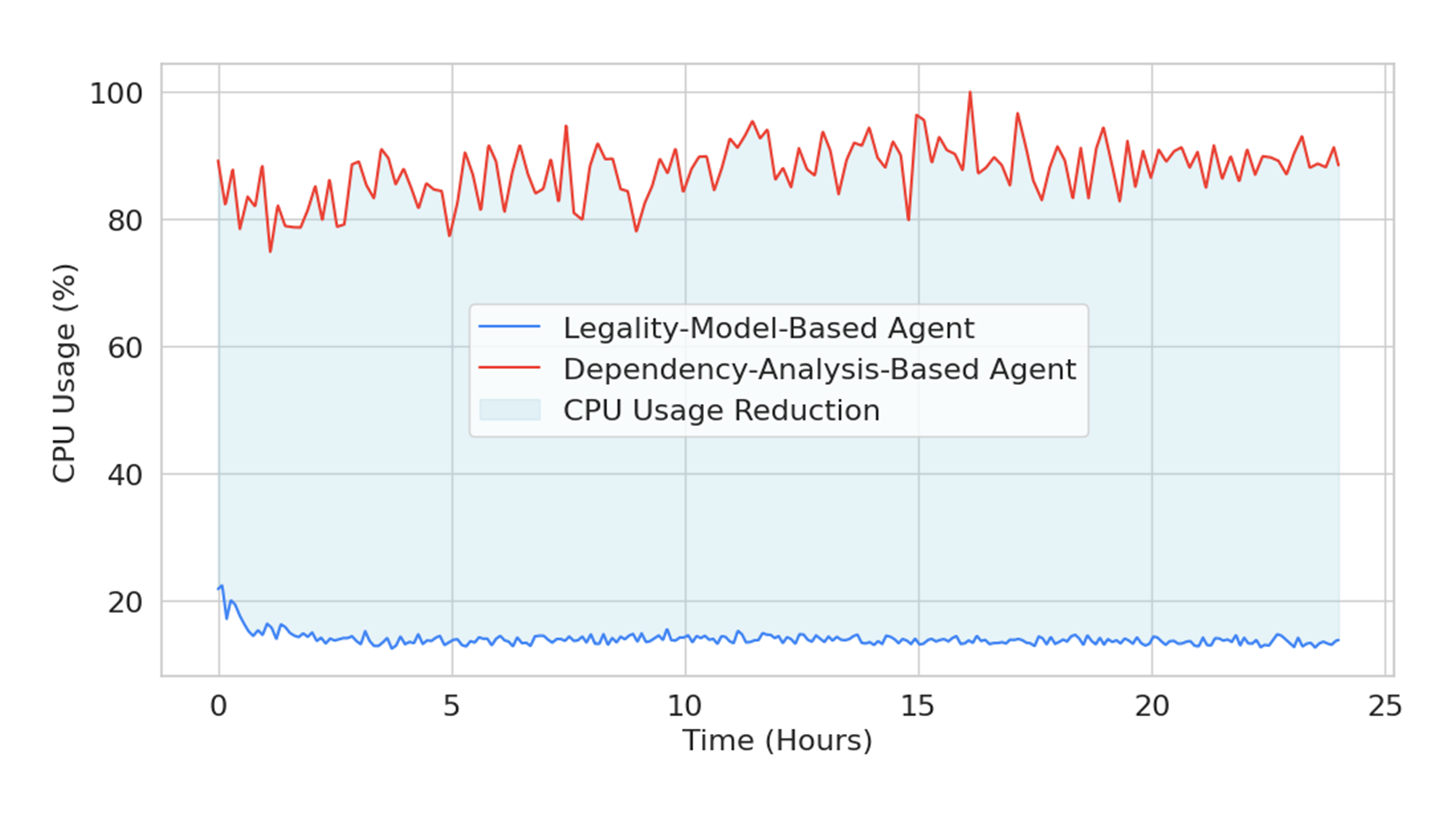}
        \captionsetup{width=0.9\textwidth}
\caption{CPU usage for the Legality-Model-Based Agent and the Dependency-Analysis-Based Agent over training time. The Legality-Model-Based Agent consistently uses significantly less CPU, indicating lower computational overhead due to faster legality checks. }
    \label{fig:CPU_usage}
	\end{minipage}
	\hfill
	\begin{minipage}[t]{0.48\textwidth}
	    \centering
        \includegraphics[width=\textwidth]{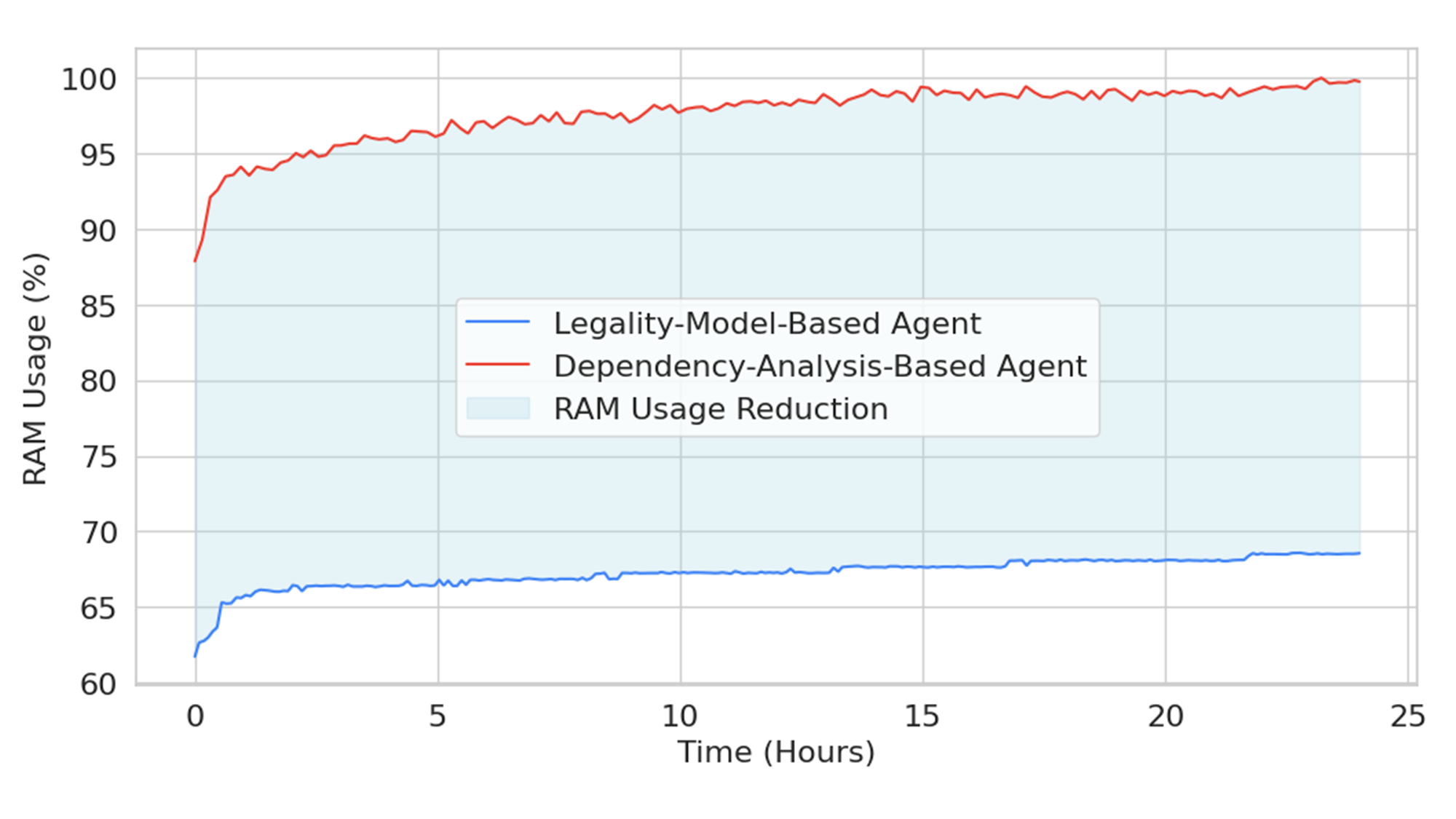}
        \captionsetup{width=0.9\textwidth}
        \caption{RAM usage for the Legality-Model-Based Agent and the Dependency-Analysis-Based Agent over training time. The Legality-Model-Based Agent demonstrates lower memory consumption, highlighting resource efficiency achieved with the proposed model. }
    
        \label{fig:ram_usage}
	\end{minipage}
\end{figure*}

Furthermore, Figures \ref{fig:CPU_usage} and \ref{fig:ram_usage} illustrate both agents' resource usage (CPU and RAM). The dependency-analysis-based agent consistently operates at near-maximum CPU and RAM usage (100\%) in both categories. In contrast, the legality-model-based agent maintains considerably lower computational demands, with CPU usage averaging below 20\% and RAM usage around 65\%-70\%. This reduced resource usage underscores the efficiency of our legality check model, which requires significantly less computational overhead, making it more suitable for resource-constrained environments.

 These results show that our model accelerates the training of the RL agent and reduces resource usage, without significantly compromising the quality of generated schedules. This is a trade-off between accuracy (reward and performance maximization) and training speed. We argue that accepting a trade-off of approximately 4\% performance degradation is justified when considering the significant advantage of achieving faster training and reducing resource usage.

\subsection{Impact of Legality Accuracy during Training on Search Efficiency and Optimization Time}
Table \ref{tab:table1} illustrates the total number of illegal transformations encountered while searching for the best schedule for each benchmark. For each benchmark, we counted all illegal transformations encountered across different input sizes(MINI, SMALL, MEDIUM, LARGE, and XLARGE). The Random-Legality-Based Agent has a poor policy that explores a substantially higher number of illegal transformations, totaling 1263 across all benchmarks. The Legality-Model-Based Agent, on the other hand, explored 884 illegal transformations, much closer to the 708 explored by the Dependency-Analysis-Based Agent. 
This difference in the number of illegal transformations directly impacts the optimization time. The Random-Legality-Based Agent requires 2.3x more time on average to find a near-optimal legal schedule than the Dependency-Analysis-Based Agent. In contrast, the Legality-Model-Based Agent needs only 1.2x more time than the Dependency-Analysis-Based Agent.

\section{Related Work}
Recently, there has been a surge in the use of ML within compilers, notably to automate code optimization~\cite{a}.

Early work in this area focused on manually extracting useful features from code~\cite{c}. These studies explored using classical ML algorithms for compiler optimization, addressing decision problems with relatively small search spaces.

Subsequent research has leveraged DL for automatic feature extraction, enabling more powerful and generalizable models. The focus has shifted towards finding the best code representation to capture the relevant information for effective optimization decisions, but always addressing limited search space problems like device mapping and code classification~\cite{hakimi2023hybrid}.
Researchers have explored using natural language processing techniques in compilers and code optimization, leveraging the similarities between source code and natural language and representing code as a sequence of tokens. The first work in this direction was DeepTune~\cite{cummins2017b}, which used an end-to-end DL model with embedding techniques to learn a distributed representation of OpenCL code.

Another direction research has explored using intermediate representations like LLVM IR  to learn better code representations that can lead to better results such as Inst2vec~\cite{ben2018neural}, and Deep LLVM~\cite{barchi2021exploration}.
Graph-based representations have also been explored, with IR2VEC~\cite{venkatakeerthy2020ir2vec} using knowledge graph embedding and by combining AST and Control Data Flow graphs ~\cite{brauckmann2020compiler}, in addition to using call graphs, control-flow graphs, and data-flow graphs to learn code representation in ProgramML~\cite{cummins2020programl}. These graph-based representations have shown state-of-the-art performance on device mapping and algorithm classification problems.

Although these works are important and propose a better representation of the code, they do not provide any way to represent the transformations with this representation, which makes them unsuitable for use in exploring a large search space of transformations because you must apply every possible sequence of transformations and compile the code to get a representation.

Another research direction explores using DL to build cost models that can guide the search for optimal compiler transformations, typically in the context of a large search space. DL-based cost models have been combined with search algorithms to explore the space of possible code transformations and identify the optimal one for the Tiramisu compiler~\cite{tiramisu21}. Similarly, Auto-TVM~\cite{chen2018learning} employs a DL cost model to predict runtime performance and optimize tensor programs for DL workloads. Another work, focusing on the Halide programming language, trains an FNN to predict the execution time of different schedules and guide the search for the best one~\cite{adams2019learning}. The authors of \cite{10.1145/3727638} raise a critical question: how can such models effectively adapt to larger or more complex programs than those seen during training? They then address this by proposing a new program representation and DL architecture.

Finally, RL has also been applied to the general problem of finding the best sequence of compiler transformations, framing it as a sequential decision-making problem~\cite{leather2020machine}. One of the earliest studies introduced a method that used RL for graph partitioning problems, aiming to optimize device placement in large computation graphs to improve performance~\cite{mirhoseini2017device}. Similarly, another study proposed a new method for the same problem using a hierarchical model to expand the scalability for larger problem sizes~\cite{mirhoseini2018hierarchical}. Researchers also proposed an RL-based automatic loop vectorization model called NeuroVectorizer~\cite{haj2020neurovectorizer}, which predicts the optimal vectorization parameters for a given loop. Additionally, LoopTune~\cite{LoopTune} introduced a deep RL method to optimize tensor computations in DL models for CPUs. Another work, AutoPhase~\cite{Huang2019AutoPhaseCP}, optimized the compiler's High-Level Synthesis phase ordering using cycle count reduction as a reward signal to guide learning. Lastly, a study using the MLGO~\cite{Trofin2021MLGOAM} framework successfully deployed RL for a production LLVM compiler, and CompilerGym~\cite{cummins2021compilergymrobustperformantcompiler} provided a standardized environment for applying RL to compiler optimization.

\section{Conclusion}
This paper proposes a novel DL model for predicting the legality of code transformations. The model is a classification model that takes the representation of full programs and their schedules as input. We evaluated the proposed model and showed that it has an F1 score of 91 on the test set. We also integrate this model with an RL agent training, achieving twice the training speed and lowering resource consumption, with minimal impact on the quality of schedules produced by the RL agent during deployment. Moreover, this model will allow us to generalize to gradient ascent methods for exploring the search space to find the best schedules for a given program while preserving its semantics. 

\section*{Acknowledgment}
This research has been partly supported by the Center for Artificial Intelligence and Robotics (CAIR) at New York University Abu Dhabi, funded by Tamkeen under the NYUAD Research Institute Award CG010. The research was carried out on the High-Performance Computing resources at New York University Abu Dhabi.

\bibliographystyle{IEEEtran}
\bibliography{refs}

\end{document}